\documentclass[epj]{svjour}
\usepackage{graphics}
\begin{document}
\title{Quark Stars: Features and Findings}
\author{Prashanth Jaikumar
}                     
%
%
\institute{Department of Physics and Astronomy, Ohio University, Athens, Ohio 45701 USA}
\date{Received: date / Revised version: date}
%
\abstract{Under extreme conditions of temperature and/or density, quarks and gluons are expected to undergo a deconfinement phase transition. While this is an ephemeral phenomenon at the ultra-relativistic heavy-ion collider (BNL-RHIC), quark matter may exist naturally in the dense interior of neutron stars. Herein, we present an appraisal of the possible phase structure of dense quark matter inside neutron stars, and the likelihood of its existence given the current status of neutron star observations. We conclude that quark matter inside neutron stars cannot be dismissed as a possibility, although recent observational evidence rules out most soft equations of state.
} 
\maketitle
\section{Introduction}
BNL-RHIC is engaged in a voyage of exploration and discovery in the high-temperature, low baryon density regime of QCD's phase diagram\cite{Harris,Nagle}. One of the central aims of this program is to characterize the deconfinement and chiral phase transition in QCD at temperatures reminiscent of the hot Big Bang. It is widely believed that a strongly interacting medium of quarks and gluons, displaying ideal liquid-like behaviour, has been created in the most energetic central Au-Au collisions at RHIC\cite{Weber,Shuryak,Peshier}. However, the lifetime of this phase is of the order of 10fm/c, requiring penetrating experimental probes that carry the imprint of the early hot partonic phase, and which are not washed out by the hadronization process\cite{Wang,Eskola,Gale}.

\vskip 0.2cm

\noindent The dense interior of neutron stars provides a complementary testing ground for quark deconfinement. The central densities inside neutron stars can be as high as 5-10$\rho_0$ ($\rho_0=2.5\times 10^{14}$ g/cc is the nuclear saturation density), and nucleons overlap to an extent that quarks and gluons become the effective degrees of freedom. Under such extreme conditions of density, it is possible that strange quark matter is energetically more stable than nuclear matter\cite{Bodmer,Witten}. If this is the case, there is a critical pressure at which a first order phase transition from nuclear to quark matter will occur. Quark matter can then comprise an arbitrary fraction of the star, from zero for a pure neutron star to one for a pure quark star, depending on the equation of state of matter at high density. 

\vskip 0.2cm

\noindent In these proceedings, we outline the rationale behind the possible stability of three flavor (up, down and strange) quark matter at high baryon density and review results from strangelet searches in heavy-ion collisions from AGS to CERN-SPS and RHIC. We consider the effects of finite size and interface energy corrections to the phase structure of stable strange matter and focus on the consequence for the surface structure of quark stars. We emphasize recent developments in neutron star observations that can shed light on the possible existence of quark matter inside neutron stars.
\label{intro}
\section{Strange quark matter in heavy-ion collisions} 
The rationale behind stable strange quark matter is the Witten hypothesis\cite{Witten}, which argues that the introduction of strangeness in up and down quark matter reduces Pauli repulsion by increasing the flavor degeneracy from up and down to up, down and strange quarks, ensuring in the process also a lower charge-to-baryon ratio for strange quark matter compared to nuclear matter. While this hypothesis is clearly not borne out for small baryon numbers, where strange baryons are definitely heavier than their non-strange counterparts, there is no observational evidence to suggest that this is also the case in bulk strange quark matter. Nuclei would not spontaneously decay to strange matter even if the latter was more stable, since that would require $\sim A$ weak reactions to occur simultaneously in a nuclear volume containing $A$ nucleons. This conversion would happen much more easily in the interior of neutron stars, where pressures and densities are supra-nuclear. The critical question regarding quark matter in neutron stars is then whether the central density of neutron stars is large enough so that strange quark matter becomes the ground state of strongly interacting matter. This question evades a precise answer because QCD is still not sufficiently well understood at neutron star densities. Lattice methods fail at such high densities due to the complexion of the measure involved in importance sampling. In the absence of concrete results from lattice studies of QCD at finite density and zero temperature, simple model-dependent studies\cite{Farhi} admit a parameter window (the parameters being the strange quark mass, the strong coupling constant and a phenomenological Bag constant) within which bulk strange quark matter is stable, even at zero pressure. If true, this implies that, if central densities inside neutron stars are large enough to create two-flavor (up and down) quark matter, or if a small nugget of cosmological/cosmic-ray origin ("strangelet") enters the star, the entire neutron matter inside the star will convert to strange quark matter by absorbing neutrons and equilibrating strangeness. 

\vskip 0.2cm

\noindent Strangelet searches in terrestrial materials, cosmic rays or as by-products of neutron star-neutron star collisions have thus far yielded negative results\cite{Mueller,Lu}. Even if strange quark matter is stable in bulk, it may be destabilized by prohibitive surface and Coulomb energy costs so that strangelets do not survive until the present day although they may have existed in the hot and dense epoch of the early universe. If so, conditions in the forward rapidity regime of ultra-relativistic nucleus-nucleus collisions (and mid-rapidity at fixed target experiments) may be able to create strangelets for a short while before they evaporate\cite{Appel,STAR}. The experimental signal searched for is a particle with a large mass-to-charge ratio that, owing to its large rigidity, would not be deflected by magnetic fields, and would be able to reach the zero-degree calorimeter (ZDC). There, it would produce a shower originating from a single point, unlike spectator neutrons which are dispersed in the transverse plane due to Fermi motion. While the strangelet search at NA52\cite{Appel} at the CERN-SPS was sensitive to long lived strangelets($\tau\sim\mu$s), the corresponding experiments at AGS\cite{VanBuren} and at RHIC\cite{STAR} were sensitive down to lifetimes of $\tau\sim$50ns. From these experiments, the production rate of strangelets was limited to less than one in $10^7-10^9$ central collisions for strangelets exceeding a mass of 30 GeV/$c^2$ and lifetime greater than a few nanoseconds.

\vskip 0.2cm

\noindent Such a low probability for producing strangelets in heavy-ion collisions is expected on theoretical grounds as well. Various models have been examined as a mechanism to produce stable strangelets (see Ref\cite{greiner} for a review). The coalescence mechanism involves forming a clump of strange matter by overlap of a sufficient number of baryons (of appropriate strangeness) with small relative momentum. This is highly improbable at collider energies. The thermal model, which has its parameters, temperature and baryochemical potential, tuned to reproduce observed particle ratios at chemical freeze-out, predicts less than one strangelet for every $10^{27}$ collisions for a strangelet with $Z/A\sim 0.1$ and mass 20GeV/$c^2$. This production mechanism falls off rapidly with increasing collider energies. The QGP distillation method relies on the enhancement of strange quarks in a QGP followed by evaporation of the baryon rich QGP through nucleons, thereby distilling strangeness and enabling a cool and stable strangelet to emerge. However, baryon rich and strangeness-rich regimes are well separated in rapidity, and there is clear evidence that the QGP cools through rapid, approximately adiabatic expansion rather than by evaporation\cite{Kolb}. Altogether, it is very unlikely that stable strangelets would be produced in a heavy-ion collision.

\label{sec:1}

\section{Strange quark matter in neutron stars}
If strange quark matter is stable only at very high baryon number ($A\gg10^{7}$), a neutron star with $10^{57}$ baryons is a natural candidate where such matter can exist. Two possibilities then arise: (i) quark matter is stable in bulk at some large value of the pressure. In this case, a first-order transition is likely to occur at some depth (density) inside the neutron star and quark matter is admixed with hadronic matter in a mixed phase whose structural details are determined by surface and Coulomb effects. (ii) quark matter is stable in bulk even at zero pressure (still at finite density, since it is self-bound). This would imply that all neutron stars are really strange quark stars, with possibly a thin layer of hadronic matter at surface. Let us examine these possibilities in more detail below:

(i) In the event that a first order phase transition occurs inside the star, a mixed phase of nuclear and quark matter can occupy a significant portion of the star's interior. This conclusion follows from the fact that there are two conserved quantities, electric charge and baryon number, which can be arranged differently in the two phases, quark and nuclear, at different equilibrium pressures. Thus, we expect a gradually increasing proportion of quark matter with increasing depth inside the neutron star. The structure and size of the rarer phase (droplets/rods/slabs) at a given density depends on the surface tension between the two phases, the curvature energy and the smallest Debye screening length. While positive surface tension and curvature energy tend to disfavor small sizes, Coulomb energy disfavors large sizes, leading to deformed structures when Debye screening effects are included. If surface tension and Coulomb costs are prohibitively large in the mixed phase, the standard picture of a sharp interface with a density discontinuity between hadronic and quark matter induced by gravity, is applicable, even though it is not the minimum energy configuration in bulk matter.

(ii) If strange quark matter is stable in bulk even at zero pressure, what we call neutron stars are really quark stars that contain quark matter almost upto the surface. At scales where the strange quark mass $m_s^2/4\mu_Q\ll 1$, with $\mu_Q$ the quark chemical potential, quark matter is effectively neutral with equal numbers of up, down and strange quarks. Near the surface of the star, however, where $m_s^2/4\mu_Q$ is not small, electrons are required to make up the deficit of strange quarks in order to form a neutral object. Microscopically, the electron distribution at the surface is governed by electrostatics on the length scale $l_e\sim 1/\sqrt{\alpha\rho_e^{1/3}}\sim 1000$fm ($\rho_e$ is the number density of electrons and $\alpha$ is the fine structure constant), while quarks are bound by the strong force (QCD scale $\sim 1$fm). Consequently, charge neutrality at the surface is impossible at scales smaller than a 1000fm in a picture where quark matter is assumed to be homogeneous. The electrons in this case distribute themselves according to the laws of electrostatics and mechanical equilibrium. They form a thin charged skin atop the star, which is held to the surface by enormous electric fields $E\sim 10^{16}$V/cm\cite{Usov}.
Such an enormous electric field is expected to emit electron-positron pairs at sufficiently high temperatures $T\geq 10^{10}$K, thereby producing a dramatic signal of hot quark stars. This signal is also transient, with a lifetime of a few days, since the star cools rapidly to lower temperatures, shutting off the pair emission\cite{Page}. Thus, observation of this signal is highly improbable from an astrophysical viewpoint.

\vskip 0.2cm

\noindent There is a more appealing alternative for the surface structure which takes into account the fact that a two-phase system can be globally rather than locally charge neutral\cite{Glendenning}. Relaxing the condition of local charge neutrality allows to reduce the strangeness fraction in quark matter at small $\mu_Q$, thereby lowering its free energy. Global charge neutrality is achieved in a mixed phase by having the phase fractions vary as a function of the pressure. This heterogeneous phase is favored when surface and Coulomb energies are negligible, as shown in the model independent approach followed in ref\cite{Jaikumar}. Since $\mu_e \ll \mu_Q$ for all reasonable equations of state describing dense quark matter, the quark pressure may be expanded in powers of $\mu_e/\mu_Q$. To second order in $\mu_e/\mu_Q$, it is given by

\begin{equation}
P(\mu_Q,\mu_e) = P_0(\mu_Q) - n_{Q}(\mu_Q) ~\mu_e + \frac{1}{2} ~\chi_{Q}(\mu_Q)~\mu_e^2 \,,
\label{eqn:Pressure}
\end{equation} 

\noindent where $n_Q(\mu_Q)=- \partial P/\partial \mu_e$ is the positive charge density, $\chi_Q(\mu_Q)=\partial^2 P/\partial \mu_e^2$ is the charge susceptibility and $P_0$ is the pressure of the electron-free quark phase. They depend on $\mu_Q$, $m_s$, and strong interactions. Typical values in the Bag model description are $n_Q = m_s^2 \mu_Q/2\pi^2 $, $\chi_Q = 2 \mu_Q^2/\pi^2$ and 
 $P_0 = 3 (\mu_Q^4-m_s^2\mu_Q^2)/4 \pi^2 - B$, where $B$ is the bag constant. 
At fixed $\mu_Q$, from Eq.~\ref{eqn:Pressure}, the quark pressure $P_q$ is zero and quark matter is positively charged when $\mu_e$ takes on the value 
\begin{equation}
\label{eqn:muecritical}
\tilde{\mu}_e=\frac{n_Q}{\chi_Q}~(1-\sqrt{1-\xi})\quad
{\rm where}~\xi=\frac{2P_0\chi_Q}{n_Q^2}\,.
\end{equation}

\noindent A mixed phase is possible when $0<\xi < 1$. In this regime, the mixed phase has lower free energy (larger pressure) than homogeneous matter. The mixed phase is however penalized by Coulomb, surface, and other finite size contributions to the energy. At zero temperature and pressure, the magnitude of the change in Gibbs free energy per quark in going from a homogeneous to a mixed phase should be more than the surface tension, i.e. the energy cost of creating a droplet surface, Then, the mixed phase is preferred over homogeneous quark matter. This critical surface tension is\cite{Alford} 
\begin{equation} 
\sigma_{\rm crit} = \frac{0.8 n_Q^2}{12\sqrt{\pi\alpha}\chi_Q^{3/2}} \ .
\label{crude_estimate}
\end{equation}

\noindent In the context of the Bag model for dense quark matter, the condition
for forming a mixed phase becomes
\begin{eqnarray}
\sigma \leq 12 ~\left( \frac{m_s}{150 ~{\rm
MeV}}\right)^3~\frac{m_s}{\mu_Q}~ {\rm MeV/fm}^2\,. 
\label{eqn:sigma_bag}
\end{eqnarray}

\begin{figure}
\resizebox{0.5\textwidth}{!}{%
\includegraphics{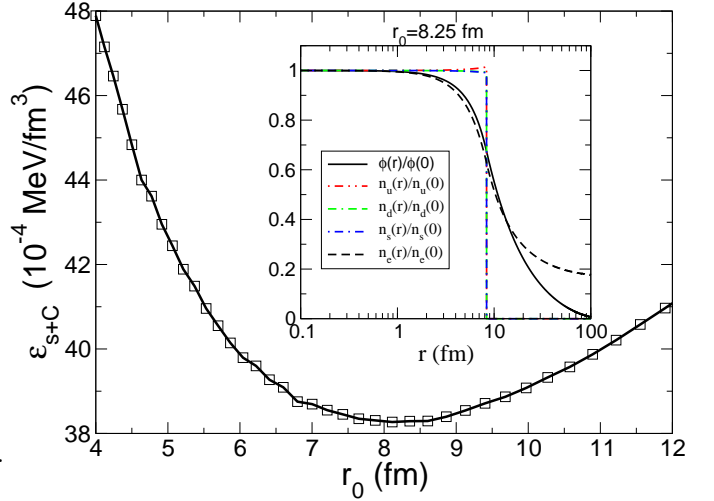}}
\caption{Surface plus Coulomb energy cost as a function of nugget size. The optimum size of the quark nugget for the choice (ii) of parameters described in the text is 8.25fm.The inset shows the quark and electron number densities, as well as the electrostatic potential $\phi$ inside the nugget.}
\label{nuggetsize}
\end{figure}

\noindent Using estimates of the surface energy of strangelets\cite{Berger,Madsen}:
(i) $\sigma \simeq 8 $ MeV/fm$^2$ for $m_s=150$ MeV and $\mu_Q\simeq 300$ MeV; and
(ii) $\sigma \simeq 5 $ MeV/fm$^2$ for $m_s=200$ MeV at $\mu_Q\simeq 300$ MeV.
The condition in Eq.~\ref{eqn:sigma_bag} implies that a homogeneous phase is marginally
favored for $m_s=150$ MeV while the structured mixed phase is favored
for $m_s=200$ MeV.  The sensitivity to $m_s$ in
Eq.~\ref{eqn:sigma_bag} and uncertainty in other finite size effects
can alter these quantitative estimates. If the structured phase is
favored, it will be composed of quark nuggets immersed in a sea of
electrons. The size of the quark nuggets in this phase is
determined by minimizing the surface, Coulomb and other finite size
contributions to the energy (see Fig.\ref{nuggetsize}).
At low temperature, this mixed phase
will be a solid with electrons contributing to the pressure while quarks contribute to the energy density - much like the mixed phase with electrons and nuclei in the crust of a conventional neutron star. This modified picture of the strange star surface has a much reduced density gradient at surface and negligible electric field unlike the old
paradigm for quark stars. In the modern viewpoint, there is no need for the
electron skin or associated large electric fields, since matter at
the surface is globally neutral. The observed luminosity from such a 
surface will be very different than from a charged surface with an electron skin\cite{JPPG}.

\vskip 0.2cm

\begin{figure}
\resizebox{0.5\textwidth}{!}{%
\includegraphics{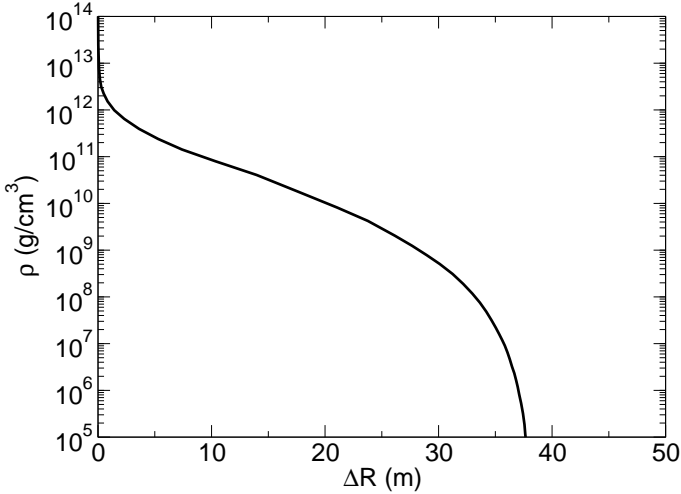}}
\caption{Density profile of the crust for a strange star with mass M=1.4 solar masses and radius R=10 km\cite{Jaikumar}.}
\label{figsurface}
\end{figure}

\noindent It is also of interest to estimate the radial extent $\Delta R$ of the mixed phase crust. This is because some neutron stars exhibit "glitches" in their rotation, when they suddenly spin-up before gradually resuming spin down. If strange stars have a large crust, where a superfluid and a lattice structure can co-exist, they could explain this glitching mechanism. Using Newtonian approximations to Hydrostatics, the size of the crust is given by\cite{Jaikumar}
\begin{equation}   
\Delta R = \frac{R}{R_s}~\frac{n_Q^2}{\chi_Q \epsilon_{\rm 0}}~R\,,
\label{eqn:analyticdr}
\end{equation}
where $R_s=2 G M/c^2\simeq 3 (M/M_\odot)$ km is the Schwarzschild radius of
the star, $R$ is the radius of the star and $\epsilon_0$ is the energy density inside a quark nugget. For $m_s=150$ MeV and $\mu_Q\simeq 300$ MeV, the Bag model with $B=65$ MeV/fm$^{3}$ yields $n_Q\simeq0.045$ fm$^{-3}$, $\chi_Q\simeq46$ MeV/fm, and $\epsilon_{\rm 0}\simeq 283$ MeV/fm$^3$. From Eq.(\ref{eqn:analyticdr}), $\Delta R\simeq 100$ meters for a star with mass $M=1.4 M_{\odot}$.  The Newtonian estimate for $\Delta R$ is close to a more accurate value for $\Delta R$ obtained by solving general relativistic equations for hydrostatic equilibrium numerically\cite{Jaikumar}. Fig.~\ref{figsurface} shows the density profile of the crust thus obtained.  
\vskip 0.2cm
\noindent The interior structure of a quark star can be quite complicated. Recently, a lot of progress has been made in understanding QCD at asymptotically high densities. In that regime, where pertubative studies are reliable, quark matter is believed to be in a color-flavor-locked (CFL) phase, characterized by quark pairing with a completely gapped spectrum. Such a phase is an electromagnetic insulator in bulk and admits no electrons, even when stressed by small quark masses\cite{Rajagopal00}. If dense quark matter indeed exists inside neutron stars, where densities are well above nuclear matter density but below the density where perturbative QCD is expected to be valid, the ground state of quark matter is uncertain. Nevertheless, in such a ``hybrid star'', attractive interactions between quarks will lead to the formation of a color superconducting state, characterized by quark pairing and superfluidity. The singlet pairing gaps can be as large as 100 MeV\cite{Rapp} and they modify transport properties due to the presence of collective excitations below the scale of the gap. At densities relevant to neutron stars, with $\mu_Q\sim 500$ MeV or less, and with the physical requirements of charge and color neutrality, the
pairing pattern can be quite complex involving phases with gapless
modes for certain quark quasiparticles. Of particular interest are the crystalline phases\cite{Bowers02}, where quarks with different Fermi surfaces pair at non-zero momentum, resulting in an inhomogenous but spatially periodic order parameter. The crystalline structure may also serve as sites for pinning rotational vortices formed in the superfluid as a result of stellar rotation, and could generate the observed glitch phenomena in neutron star spin-down.
The gapless and crystalline phases can also lead to temperature dependences that
are modified from the usual forms in ungapped quark matter. These
phases also have unique dispersion relations for certain quark
quasiparticles, and consequently, a specific heat per unit volume that
is also different from ungapped quark matter. These two factors imply
a change in the stellar cooling curve that can be confronted by
observations\cite{Jotwani}. The question as to which is the preferred stable state of quark matter at intermediate densities and physical strange quark mass remains an open one at this time.
 
\label{sec:3}
\section{Constraints from neutron star observations} 

Neutron star observations can help in constraining the equation of state of dense matter, and also in distinguishing between different models for the crust as discussed above. Individual neutron star masses are most precisely determined by measuring post-Keplerian orbital parameters in close binary systems. Neutron star masses thus determined lie in the range 1.18-1.44 $M_\odot$ and have errors of less than a tenth of a percent. If the radius can also be ascertained to high accuracy, the equation of state of dense neutron star matter can be pinned down. In practice, there are several complications that make radius measurements a challenge so that it is only possible to infer the radius at infinity which is related to the true radius of the star through the relation $R_\infty=(1+z)~R$ with the red-shift factor $(1+z)^{-1}=\sqrt{1-2GM/Rc^2}$. Consequently, instead of measuring a radius we can only infer a relation between mass and radius. It has recently been realized that compact objects in low-mass X-ray binaries (LMXBs) which exhibit  X-ray bursting behavior may provide a promising new avenue to determine, simultaneously, both the mass and radius of a neutron star\cite{Ozel}, thereby setting firm constraints on the equation of state of dense matter. In these objects, there is the potential to observe, in addition to the quiescent luminosity that can be used to infer $R\infty$, the Eddington luminosity during the burst and the gravitational red-shift through direct observation of the shift in identifiable atomic absorption lines in the atmosphere. The simultaneous determination of mass (2.10$\pm$ 0.28~$M_{\odot}$) and radius (13.8$\pm$1.8 km)\cite{Ozel} of the bursting LMXB EXO 0748-676 eliminates most soft equations of state and is compatible only with the stiffest neutron or quark matter equations of state. It does not rule out all quark matter equations of state, therefore, hybrid stars or strange stars remain viable\cite{JSB}. However, this and other heavy neutron star candidates and the rather large inferred radius for EXO 0748-676 disfavors the scenario in which significant softening due to a phase transition at high density occurs. These recent developments are reviewed in ~\cite{Redpage,JRS2}.

\section{Conclusions}
The existence of stable strange quark matter remains a difficult proposition to
veto or verify. The true ground state of strongly interacting matter at high density is as yet unknown, but useful constraints are placed by recent neutron star observations. In addition, observed astrophysical phenomena such as glitches, quasi-periodic oscillations in accreting neutron stars, thermal radiation from
quiescent LMXBs and seismic vibrations during magnetar flares
can potentially yield valuable information about the neutron star crust and
interior, thereby constraining the equation of state at high density. It is apparent that recent neutron star observations disfavor the appearance of soft exotic (non-hadronic) matter at high density. They do not as yet completely rule out all plausible quark matter equations of state. Given that strangelet searches in astrophysical as well as terrestrial (heavy-ion collider) experiments have yielded null results, it appears that our access to the underlying degrees of freedom in QCD may be limited to the transient partonic fireball created in ultra-relativistic heavy-ion collisions. 

\vskip 0.2cm
\noindent I acknowledge beneficial discussions with Sanjay Reddy, Andrew Steiner, Rachid Ouyed, Andreas Schmitt, Mark Alford and Giorgio Torrieri; and I thank the organizers of Hot Quarks 2006 for an invigorating workshop. This work is supported by US DOE grant
DE-FG02-93ER40756.

%

\end{document}